\documentclass[11pt,fleqn]{article}
\usepackage{graphicx}
\usepackage{amsfonts}
\usepackage{amstext}
\usepackage{amsmath}
\usepackage{amsbsy}
\usepackage{amssymb}
\usepackage{multicol}
\begin{document}
\thispagestyle{empty}
\sf
\vspace{-5mm}
\centerline{\Huge Neutrinos as astrophysical probes}
\vspace{3mm}
\centerline{\large Flavio Cavanna$^a$, Maria Laura Costantini$^{a}$,
Ornella Palamara$^b$, Francesco Vissani$^b$}
\vspace{2mm}
\centerline{\small\em $^a$Universit\`a dell'Aquila e INFN -  
Via Vetoio, I-67010 L'Aquila, Italia}
\centerline{\small\em $^b$INFN, Laboratori Nazionali del Gran Sasso -  
S.s.~17 bis, I-67010 Assergi (AQ), Italia}
\vspace{1mm}
\begin{quote}
\small
The aim of these notes is to provide a brief review of the 
topic of neutrino astronomy and in particular 
of neutrinos from core collapse supernovae.
They are addressed to 
a curious reader, beginning to work in  a 
multidisciplinary area that involves 
experimental neutrino physics, astrophysics,  nuclear physics and
particle physics phenomenology.
After an introduction to the
methods and goals of neutrinos astronomy, we focus
on core collapse supernovae, as (one of) the most promising 
astrophysical source of neutrinos.
The first part is organized almost as a tale, 
the last part is a bit more technical.
We discuss the impact of flavor oscillations on the supernova neutrino 
signal (=the change of perspective due to recent achievements) and 
consider one specific example of signal in detail.
This shows that effects of oscillations are important,  
but astrophysical uncertainties should 
be thought as an essential systematics 
for a correct interpretation of future experimental data.
Three appendices corroborate the text with further 
details and some basics on flavor oscillations;
but no attempt of a complete bibliographical survey is done
(in practice, we selected a few references 
that we believe are useful for a `modern' introduction 
to the subject and suggest the
use of public databases for papers \cite{web1} and 
for experiments \cite{web2} to get a more  
complete information).
\end{quote}
\small
\tableofcontents
\newpage
\section{Neutrino astronomy, methods and goals\label{sec:nuastr}}
\subsection{Main neutrino features\label{ssec:nufeat}}
Neutrinos (and anti-neutrinos) of   electron-, muon- and tau-flavor, 
are stable, neutral particles.
This makes them important astrophysical probes;
they are expected 
to point in the direction of the astrophysical site 
of production, as in the more standard case of 
astronomy with photons.\footnote{Protons 
and nuclei of cosmic ray radiation, instead, 
are deflected by galactic of $\sim$ few $\mu$G
and extragalactic magnetic fields, at or below nG.
They are not expected to point to their sources except perhaps at 
the very highest energies. Fast galactic neutrons instead 
are another interesting neutral probe.}
Here we have in mind the case of
`point astrophysical sources'; but of course
`diffuse sources' are also of importance.

In normal conditions, neutrinos are invisible. 
However, they can sometime interact and  carry away 
or deposit energy in terrestrial detectors. By contrast, 
photons are much more easily absorbed than neutrinos;
they can be observed more easily, but for the same reason
their propagation can be more easily affected.
In certain cases, neutrinos will be the most important 
signal (think for instance to 
neutrinos from big-bang nucleosynthesis, 
from the sun, or from a core collapse supernova).

Some neutrino interactions are of special 
interest for the following discussion. First, 
\begin{equation}
\nu_e + e\to \nu_e+ e,\ \ \ \nu_\mu +e\to \nu_\mu+ e\ \ 
\mbox{[CC and NC elastic scattering]}
\end{equation}
In these reactions an $e$ at 
rest -- say, from an atom -- is hit by the 
neutrino and acquires kinetic energy. 
An important feature is that the hit $e$ maintains the direction 
of the neutrino when the $\nu$ energy $E_\nu\gg m_e$ (``directionality'').
The cross-section is low, $\sigma_\nu\sim G_F^2 m_e E_\nu$
($G_F$ is the Fermi coupling). \\
The (lowest energy) neutrino reactions are those of absorption 
on nucleons  and on nuclei:
\begin{equation}
\begin{array}{l}
\bar{\nu}_e +p\to e^+ +n,\ \ \ {\nu}_e +n\to e+ p \\
\bar{\nu}_e +(A,Z) \to e^+ +(A,Z\!-\!1),\ \ \ {\nu}_e +(A,Z)\to e+ (A,Z\!+\!1)
\end{array}
\end{equation}
these reactions  have usually a threshold, 
and are only slightly directional (more 
quantitative statements requires care 
to details, see e.g., App.~\ref{xsec}).
NC cross-sections on nuclei can be as  large as
$\sigma_\nu \sim G_F^2 A^2 E_\nu^2$, when a 
nucleus composed by $A$ nucleons reacts 
as a whole (coherent scattering).
At higher energies, the absorption 
cross sections on nuclei 
become $\sim G_F^2 A~ m_p E_\nu$ 
(incoherent scattering). In this case, 
the nucleus is broken and/or hadronic 
resonances are excited.

\subsection{Concepts of neutrino telescopes\label{ssec:nutel}}
Let us describe some concepts of neutrino detector, to illustrate 
what people mean by a `neutrino telescope'.\footnote{Warning: As it is 
common in physics, different concepts are blurred and useful
at best for orientation; 
in present case, they depend
on the type of particle, on 
the size of the detector ...}
(Supernova neutrino detectors fall 
in the first concept, normally.)

\noindent $\star$
{\em One can instrument a large volume, possibly  vetoing
for external particle and wait for a charged particle coming 
apparently from nowhere---in actuality, created 
by a neutrino interaction. (Better to be underground for
low counting rates, like those related to natural 
neutrino radiation.)}
Active volume can be  
a scintillator, a \v{C}erenkov radiator, 
a layered target, a `bubble chamber-like' detector. This method
works from sub-MeV to several GeV energies, because it is subject to 
the condition that the (main part of the) 
event is contained in the detector.
The number of events
scales as
$$N=\mbox{Number of targets}\times \sigma_\nu\times 
\Phi_\nu\times \mbox{Time}  ~~~~\mbox{ ($\Phi$ denotes generically a flux).}$$
In particular, the event rate scales as the \underline{volume} 
of the detector.

\noindent $\star$
{\em One can set a muon counter and timing system
underground (or underwater or under-ice), 
for muons that originate from neutrinos -- as those coming from below}. 
Detectors are located underground to avoid cosmic ray muons.
This is the oldest method and works since
muons suffer of mild energy losses till $\sim 500$ GeV (that 
corresponds roughly to Range$_\mu\sim 1$ km in water).
It applies from energies around a GeV till several hundred TeV;
then the earth becomes opaque even to neutrinos (see e.g., \cite{teresa}).
The number of events and the $\nu$-induced muon flux
scale respectively as:
$$N=\mbox{Surface}\times \Phi_\mu\times \mbox{Time}~~~\mbox{ where }~~
\Phi_\mu=\Phi_{\nu_\mu}\times \sigma_{\nu_\mu}\times N_A\times 
\mbox{Range}_\mu.$$
In particular, the signal scales as the \underline{area} 
of the detector (actual target being the earth, the water or 
the ice where the detector is located).

\noindent $\star$
{\em By an extension of previous concept, one could use the 
earth atmosphere as a target for 
high energy neutrinos to produce inclined air showers;
or, use mountains to convert almost horizontal  $\nu_\tau$ of very 
high energy into visible tau's}.
In this way, we could observe neutrinos of highest energies.
The search of inclined air showers is just a spin-off of 
extensive air shower arrays research activity. 
Till now, however, no positive detection has been claimed.

\vskip3mm
In principle we would like to {measure} 
a lot of quantities:
a)~direction of the charged lepton;
b)~its energy;
c)~its charge;
d)~tag the flavor;
e)~tag the time of arrival;
f)~check occurrence of secondaries ($n,\gamma,$ charged hadrons).
In practice, one has to find a compromise between 
the various and contrasting needs of an experiment, 
e.g.~between the wish to have a very `granular'  detector able to see all  
the details of the reaction and the need to monitor a big amount of matter.

\subsection{Chances for neutrino astronomy\label{ssec:nuastr}}
In short, the goal is to use neutrinos to probe astrophysical
sources; the information from $\nu$ can be complementary 
to the one from $\gamma$. Some important 
possibilities in this connection are:
\vskip3mm

\noindent {\em (1)} \underline{Solar neutrinos} [0.1-20 MeV] 
There is little doubt that this is `$\nu$-astronomy'.
Among the results of a very successful 
program of  observations pioneered by 
Homestake we quote:\footnote{We should recall 
the important role of certain theorists: 
J.N.~Bahcall, whose activity has been very 
supportive to Homestake since the beginning
and G.T.~Zatsepin and V.A.~Kuzmin, who strongly advocated 
the importance of solar neutrino astronomy.}
a) low energy $\nu$ experiments Gallex/GNO and SAGE
prove that the pp-chain (initiated by $pp\to D e^+\nu_e$)
is the main energy source; 
b) the physics of the center of the sun
($\rho_c \sim 150$ g/cm$^3$) is probed.
There is consistency with the theory of solar 
oscillation eigen-modes (helioseismology). c) Neutrino
oscillations 
of a type predicted in MSW  
theory
are indicated.\footnote{This reconciles 
SNO observations (1/3 of expected $\nu_e$) with those of  
low energy $\nu$ experiments (where 
the deficit is less than 1/2). 
KamLAND experiment supports strongly
this picture; more discussion later.} 
Future observations will aim at the Beryllium
line (Borexino, KamLAND) and at real time pp-neutrino detection.\\[2ex]
{\em (2)} \underline{Atmospheric neutrinos} [0.05-1000 GeV]
primary cosmic rays (CR) come isotropically on earth atmosphere and 
they are not completely understood;
they are not thought as astronomy, but
they belong to astrophysics as much as to particle physics.
Atmospheric neutrinos
give a very significant indication of oscillations,
especially thanks to Super-Kamiokande 
results.\footnote{MACRO, Soudan2 and K2K support these results. 
Again K2K, Minos and CNGS long-baseline experiments will
further test these results with man-made neutrino beams.} 
The study of CR secondaries as the electromagnetic component, 
muons or atmospheric neutrinos, 
permits us to investigate CR spectra and their interactions with 
earth atmosphere (which is not that different from possible sites of 
production of CR).
In the present context, atmospheric neutrinos will be 
thought just as an important background.\\[2ex]
{\em (3)} \underline{Neutrinos from cosmic sources} [unknown energies]
This is a vast field and includes a large variety 
of approaches of observation and of objects; 
presumably, also unknown objects \cite{halz}.
For instance, one can search for an excess of neutrino
events over the expected background 
by selecting a solid angle--observation window--around a cosmic 
source (say, an active galactic nucleus) 
or an appropriate time window around a cosmic events 
(say, a gamma ray burst). Other possibilities
are to search for self-trigger (excess of multiple `neutrino' 
events), or coincidence with other neutrino- or with 
gravitational wave-detectors.
The observation of point (or diffuse) sources 
is a very important goal: e.g., $\nu$ (and $\gamma$) astronomy 
above TeV can shed light on the problem of the origin of CR.
Till know, several experiments like LSD, MACRO, LVD, 
Super-Kamiokande, Soudan2, Baksan, AMA\-NDA, EAS-TOP, HiRES
and other ones produced upper limits on the fluxes. 
In future, this type of search will 
be conducted by  ANTARES, AUGER, ICECUBE. One of the main 
hopes is that the neutrino energy spectrum remains very hard 
till $\sim $100 TeV, as suggested by observed gamma spectra
at 1-10 TeV (another one is that the prompt neutrino 
background--from charm--is not overwhelming.)\\[2ex]
{\em (4)} \underline{Supernova neutrinos} [few-100 MeV]
(this `cosmic' source is singled out, since it is 
the topic of the rest of the paper). 
As  recalled in next section, most of core collapse
supernova energy is carried off by neutrinos of all flavors.
About  20 events were detected in 1987 by simultaneous 
observations\footnote{Five other events have been
detected by LSD experiment about five hours before the main signal,
see V.L.~Dadykin {\em et al.}, JETP Lett.~45 (1987) 593. 
Recently, it was remarked that they could be 
explained postulating a pre-collapse phase of emission
where only non-thermal $\nu_e$ of $\sim 40$~MeV are emitted:
see V.S.~Imshennik and O.G.~Ryazhskaya, ``Rotating collapsar
and a possible interpretation of the LSD neutrino signal from SN1987A'',
to appear in print (preliminary reports presented at `Markov Readings',
INR, May 2003, Moscow and LNGS Seminar Series, Sept.\ 2003, L'Aquila).
In this hypothetical phase of emission 
called also `cold collapse' the 200 tons of iron surrounding 
the LSD detector were the most effective target of 
terrestrial detectors.}
of Kamiokande~II, IMB, Baksan detectors \cite{sn1987a} from such 
a supernova, SN1987A,
located in the Large Magellanic Cloud, at a distance 
$D\approx 50$ kpc. Usually,
all these events are attributed to 
inverse $\beta$-decay, the one with the 
largest cross-section 
(see App.~\ref{xsec}).
The experimental detection of these events
begun extragalactic neutrino astronomy.
The agreement with the expectations is reasonable.\\
Many operating neutrino detectors like Super-Kamiokande, SNO, 
LVD, KamLAND, Baksan, AMANDA could be blessed by 
the next galactic supernova. 
Other detectors like ICARUS and Borexino will 
also be able to contribute to 
galactic supernovae monitoring in the future.
This activity will have a big 
payoff in astro/physics  currency: core collapse SN 
are a source of infrared, visible, $X$, 
and $\gamma$ radiation and possibly of gravitational waves;
they are of key importance for origin of galactic CR,
for reprocessing of elements, presumably for the dynamics 
of magnetic fields; they are likely to be related to 
cosmic phenomena like gamma ray bursts; etc. 
In the following, we focus only on supernova neutrinos. 

\subsection{Galactic, extragalactic and relic supernovae}
We close this introduction by classifying
and discussing the possible observations of SN neutrinos.
(Note that, unless said otherwise, 
the term supernova means always 
{\em core collapse supernova} in these notes, 
even though this is an abuse of notation -- supernovae
of type Ia are very important in cosmology and astrophysics, 
and are {\em not} core collapse events). 

The hope of existing neutrino telescopes is the explosion of a 
{\bf galactic supernova}, for the simple fact that the 
$1/D^2$ scaling of the flux is severe. In water or scintillator detectors
one expects roughly 300 
$\bar\nu_e$-events/kton, for a distance $D$=10 kpc --
when our galaxy has a radius of some 15~kpc and 
we are located at 8.5 kpc from its center.\footnote{One 
could expect that the chances of getting a supernova 
where matter is more abundant are higher
(the galactic center), but
one can also object that younger matter, conducive to SN formation,
lies elsewhere (in the spiral arms). 
However, we are unaware of the existence of
a `catalog of explosive stars of our galaxy', or of calculations 
of weighted matter distributions of our galaxy.}
Various authors estimated the rate of occurrence of
core collapse supernovae;
for our galaxy, this ranges from $\sim 1/(10$ y) to $\sim 1/(100$ y). 
A recent study \cite{turatto}
of the correlations of $\sim 200$ 
observed supernovae at cosmological distances with
the blue luminosity of their host galaxy 
yields  1/(50-100 y).\footnote{The 
main unknown comes from the fact 
that we ignore which is the type of the galaxy 
that guests us; this implies the factor 2 
of uncertainty.}
A $\sim 1/(10$ y) lower limit can be 
already established, since existing $\nu$-telescopes 
did not observe any event yet. 
Often, one recalls 
the possibility that SN events can take place 
in optically obscured regions of our galaxy;
however, one should also remind that, beside $\nu$'s, there 
are other manners to investigate the occurrence of such 
a phenomenon, e.g., from the released infrared radiation.

Curiously enough, galactic neutrino astronomy is still to begin,
but as recalled 
{\bf extragalactic neutrino} astronomy begun several years ago with
SN1987A. In principle, one should profit of the 
wealth of galaxies around us (say, those in the `local group') to 
get events at human-scale pace. 
In practice this is difficult, because 
core collapse SN takes place only in
spiral or irregular galaxies and not in 
elliptical ones.\footnote{Their stellar population is 
older and star forming regions 
are absent or very rare; in a sense, the stars 
of 10-40 solar masses are a problem of youth.} 
The only other large spiral galaxy of the local group 
is Andromeda (M31) but {\em (1)}~its mass is presumably 
half of our galaxy, {\em (2)}~its distance is about 700 kpc.
A half-a-megaton detector (as the one suggested 
as a followup of Super-Kamiokande to continue 
proton decay search) should get 30 events 
if efficiency is unit.
Perhaps, the best chance would be another SN 
from Large Magellanic cloud (an irregular galaxy) 
but the odds for such an event are not high.

Another interesting possibility is the 
search for {\bf relic supernovae}, namely  
the neutrino radiation emitted from past supernovae. 
The practical method is to select an energy window around 
$20-40$~MeV,  where atmospheric or other neutrino 
background is small, searching for an accumulation of neutrino 
events there with more-or-less known distribution. 
The best limit has been obtained by the Super-Kamiokande water-\v{C}erenkov 
experiment \cite{malek}, 
and the sensitivity is approaching
the one requested to probe interesting theoretical models. In principle,
one can suppress the main background (muons produced 
below the \v{C}erenkov threshold) by identifying the neutron from 
neutrino inverse $\beta$-decay reaction.
This could be perhaps possible 
by loading the water with an appropriate nucleus with high n-capture rate, 
that should absorb the neutron and yield visible $\gamma$ 
eventually see e.g., \cite{jb}.\footnote{Neutron identification by
$p+ n\to D+\gamma$ (2.2 MeV)
was proved in scintillators 
(furthermore, no \v{C}erenkov threshold 
impedes); however no existing scintillator 
has a mass above 1~kton.}
 
\section{Supernova neutrinos\label{sec:SNnu}}
In Sec.\ref{ssec:gc} and \ref{ssec:nf} we present theoretical 
expectations on supernova neutrinos.
More precisely, we describe 
the expected sequence of events of the
`delayed scenario'.
This is the current theoretical framework \cite{delayed} \cite{janka}, 
possibly leading to SN explosion.
In Sec.\ref{ssec:no} we discuss generalities of 
SN neutrino oscillations. We provide the basic concepts and formulae
and discuss the impact on the fluxes. 
(The basic terminology and results are recalled 
in App.~\ref{o}, but a real beginner could conveniently 
consult review articles or texts before reading this section.
For a more advanced reading, we list in 
Ref.~\cite{alyosha} some recent research 
works on oscillations of supernova neutrinos.) 
Finally, we complete the discussion and 
show an application of the formalism 
in Sec.\ref{ssec:eos}, by considering the 
reaction $\nu_e$ Ar$\to $K$^{*} e^-$
as a signal of supernova neutrinos in an Argon based detector.

However, the reader should be warned: 
at present it turns out to be difficult (perhaps 
impossible) to simulate a SN explosion.
This could be due to a very complex dynamics; 
or, it could indicate that 
some ingredient is missing 
(such as an essential role 
of rotation, of magnetic fields, etc);
or that there is nothing like a `standard explosion';
or, worse, a combination 
of previous possibilities.
In short, we have not a `standard SN model' yet
and this makes supernova neutrinos even more interesting.

\subsection{Gravitational collapse and the `delayed scenario'\label{ssec:gc}}
Usually, the life of a star is characterized by 
a quasi-equilibrium state between gravity and nuclear forces.
However, the dramatic conclusion the brief-some million years-life  
of a very massive star of $\sim 10-40\ M_\odot$ is 
something very different, a core collapse supernova.

Stellar evolution forms an iron-core, inert to nuclear reactions.
This is supported by degeneracy pressure of (quasi)free electrons,
but when it exceeds the Chandrasekhar mass of 
$\sim 1.4 M_{\odot} $ (radius $\sim 3 \times 10^{3} $ km)
it collapses under its weight.\footnote{The gravitational
pressure is $P_g\sim G_N M^2/R^4$.
The $e^-$ pressure is $P_e\sim u/v_e$ 
where $v_e=1/N_e$ is the specific volume
and the internal energy $u$ is 
$c p_F$ or $p_F^2/(2 m_e)$ 
depending on whether electrons 
are relativistic or not
($p_F$=Fermi momentum); thus, 
$P_e\sim \hbar c~N_e^{4/3}$ or $P_e\sim \hbar^2/(2 m_e)~N_e^{5/3}$.
Since the electron density $N_e\sim M/(R^3 m_n)$,
non-relativistic $e^-$ lead to the scaling $P_e\sim 1/R^5$
and an equilibrium can be reached; for relativistic ones 
$P_e\sim 1/R^4$ and equilibrium is impossible 
after the core reaches the Chandrasekhar 
mass of $M\sim (\hbar c/G_N)^{3/2}/m_n^2$.}
The neutron density of the 
innermost part of the core (the `inner-core', $\sim 0.6 M_{\odot} $)
enlarges progressively due to iron photo-dissociation 
followed by electron capture -- ``infall'' phase.
When it reaches nuclear densities 
the increase in matter pressure is sufficient to halt the collapse.
The `outer-core' (which is still free-falling 
onto the center of the star) undergoes a bounce on
the stiff inner-core. 
In this moment, an outward-going shock-wave forms, producing a 
prompt neutronization 
in the shocked material whose mass is about $\sim 0.4 M_\odot$
-- ``flash'' phase. 
Then, the shock wave enters a phase of stall, 
trying to make its way through
the outer part of the core. This  turns the propagating
wave into an shock of accretion 
that involves 
rest of the initial iron core, $\sim 0.5 M_\odot$
-- ``accretion'' phase.
During this phase, convective motions 
and neutrinos (the `delayed mechanism') {\em should} revive the shock
(that subsequently will eject outer star's layers -- the SN explosion).
The inner core settles in a new quasi-equilibrium state 
called protoneutron star,  that smoothly cools and contracts radiating 
neutrinos of all types -- ``cooling'' phase.
Eventually this leads to the formation of a neutron 
star (NS),  occasionally seen as a pulsar. 
Its mass is $M_\mathrm{ns} = 1-2\ M_{\odot} $, 
and its radius scales roughly as 
$R_\mathrm{ns} \approx 20\mbox{ km} \times(M_\odot/M_\mathrm{ns})^{1/3}$, 
due to degenerate character of the equation of state.
The main features of the collapse process, subdivided in the various phases
mentioned above, are summarized in Tab.~\ref{tab:SNcoll}.

\begin{table}[t]
\caption{\footnotesize\textsf{\textit{Schematic description of 
collapse and neutrino emission in the delayed scenario.
(SN progenitor mass $M_\star\sim 13\pm 3 M_\odot$).
In the first four rows, the main phases are identified. 
Their conventional names  are given in Column-1 and the expected 
dynamics is described in Column-2.
Only the main reactions and $\nu$-processes (Column-3) are listed.
The last row  refers to the newborn n-star: when temperature decreases 
to $\sim 10^9$~K it becomes transparent to neutrinos; the emission 
continues for $\Delta t \sim 10^5$ yr, until the
temperature drops to $\sim 10^8$~K.\label{tab:SNcoll}}}}
{\footnotesize
\begin{center}
\begin{tabular}{|c|c|c|c|c|}
\hline\hline
{\sf Collapse Phase}          & {\sf Dynamics }                  & {\sf $\nu$ Process}       & {\sf Duration}             & {\sf Energetics}   \\
\hline\hline
``Infall"                           & Iron Core Collapse  &   $\nu_e$-emission   & $\sim 100$ ms   &                    \\
(early neutronization) &                                     &    $e^-+p\to n+\nu_e$   &                              &                    \\
 of inner core                & $\gamma +Fe\to 13^4He+4n$  &                   & $\Delta t_{inf}\lesssim 25$ ms &   $\delta_{inf} \le 1$\%~${\cal E}_B$    \\ 
 $\sim 0.6 M_\odot$     &   $^4He\to 2n+2p$  &  $\nu_e$-trapping     &                              &                                                 \\
                                        &                                    &   $\nu_e+A\to \nu_e+A$  &                       &                                                 \\
                                        &                                    &  $\nu_e+n\to p+e^-$     &                              &                                                 \\
\hline
``Flash"                         & Bounce. Shock wave  & $\nu_e$-burst         & [$t\equiv t_0$]      &                                                 \\
(prompt neutronization) &				  & $e^-+p\to n+\nu_e$        & 			   &  $\delta_{fl}\sim 1$\%~${\cal E}_B$     \\
of (part of ) outer core &$\gamma +Fe\to 13^4He+4n$ & at $\nu_e$-sphere & $\Delta t_{fl}\lesssim 10$ ms &                 \\
$\sim 0.4 M_\odot$     &                                     & 				          &			     &			       		 	   \\
\hline
``Accretion"                  &  Stall of  shock wave & $\bar\nu_e$-emission  &                          &                                                   \\
Mantle neutronization &  				     & $e^+ +n\to p + \bar\nu_e$ &  		     &  					 \\
$\sim 0.5 M_\odot$     &  $\gamma\to e^+e^-$ & 				          &				   &				  \\
         				  &                                     & $\nu_i$-emission      &			     &			       		 	   \\
			           &    				      & $e^+e^-\to \nu_i \bar{\nu}_i$ &$\Delta t_{accr}\lesssim 500$ ms& $\delta_{accr}\approx 10$ \%~${\cal E}_B$ \\
			          & Proto n-star formation& 				          &			     &			       		 	   \\
Delayed shock revival & 			           &   $\nu$-heating			 &				&					\\
 	 		         & SN explosion    & $\nu_e+n\rightleftharpoons p+e^-$  &			     &			       		 	   \\
	 		         & 				   & $\bar\nu_e + p \rightleftharpoons n+e^+  $         &			     &			       		 	   \\
			         \hline
``Cooling"                    & Mantle contraction    & $\nu_i$-emission   & 				   & 						   \\
residual neutronization &  $\gamma\to e^+e^-$ &$e^+e^-\to \nu_i \bar{\nu}_i$& $\Delta t_{cool}\sim 10$ s& $\delta_{cool}\sim 90$ \%~${\cal E}_B$ \\
			          &			 &at $\nu_i$-sphere&			     &			       		 	   \\
\hline\hline
n-star 		         & 				    & 	$\nu$-`fading'	     &			     &			       		 	   \\
$M_{\mathrm{ns}}\sim 1.4 M_\odot$ &  Steady state  & $n~n\to n~p~e^-\bar\nu_e$&      			 & few \%~${\cal E}_B$   \\
$R_{\mathrm{ns}}\simeq 18$ km       &    			       &$n~p~e^-\to n~n~\nu_e$ &			     &			       		 	   \\
$\rho\simeq3\times10^{14}$ g/cm$^3$   &                & 			    &			     &			       		 	   \\
\hline\hline
\end{tabular}
\end{center}
}
\vskip-5mm
\end{table}
The most important aspect to note is that the 
gravitational binding energy released during the collapse process 
(up to the n-star formation) is huge, about 
\begin{equation}
{\cal E}_B \simeq G_N \
\frac{3}{5}\frac{M^2_\mathrm{ns}}{R_\mathrm{ns}}
\approx (1-5) \times 10^{53}\ \textrm{erg}
\label{eq:eb} 
\end{equation}
(3/5 is for a uniform density distribution)
that is about $\sim 10 \%$ of the n-star rest mass 
energy $M_\mathrm{ns}\, c^2$.
This is much bigger than the kinetic energy of the ejecta  
$E_\mathrm{kin} \sim 10^{51} $ ergs $ \approx 1 \% $ ${\cal E}_B $
(a typical velocity of the shock wave is 4-5000 
km/s, ejecta mass $M_\mathrm{ej} \sim 10M_\odot$).
Also much bigger than what is needed 
to dissociate the outer iron core 
$0.6 M_\odot/m_n \times (2.2\ \mathrm{MeV}) =2\times 10^{51}$ erg
since the mass of ${}^{56}$Fe is 123~MeV smaller 
than $13 m_\alpha +4 m_n$ -- but this could be 
optimistic and the energy losses 
suffered by the shock wave even larger.
The energy that goes in photons is very small, 
$E_\mathrm{lum} \approx 10^{49}$ erg 
$ \approx 0.01 \% $ ${\cal E}_B $
(sufficient to outshine host galaxy though!) and 
the gravitational wave part is unknown (and depends on the 
detailed dynamics of the collapse) 
but it is probably even less.\footnote{A naive 
guess is $G_N (M v^2/2)^2/R\sim {\cal E}_B\beta^4$; 
it means some billionth of ${\cal E}_B$
with $v\sim 4000$ km/s.}
The overwhelming part of this huge energy is carried away by 
neutrinos (main reactions leading to $\nu$ production in the various 
collapse phases are reported in Tab.~\ref{tab:SNcoll}). 
The neutrino `luminosity' can be roughly 
estimated noting that $ \sim 10^{53} $ erg are emitted in 
a few seconds in the cooling timescale, 
and thus ${L}_{\nu} \approx 3 \times 10^{19} \ {L}_{\odot}$:
the supernova neutrino burst outshines the 
entire visible universe. (Incidentally, we feel 
there is something poetic in these quasi-spherical
SN neutrino shells that propagate freely 
in the Universe).

\subsection{Neutrino fluxes\label{ssec:nf}}
Here, we describe in some detail the
neutrino fluxes. First we discuss the general 
characteristics and present a phenomenological survey,  
and then we discuss how their 
luminosity, energy spectrum and  possible 
non-thermal effects can be parameterized.
We ought to recall the three relevant types of neutrino fluxes:

\centerline{\fbox{$\nu_e,\ \  \bar{\nu}_e\ \mbox{  and  }\ \nu_x$}}

\noindent where $x$ is anyone among muon and tau (anti)neutrinos. In fact,
$\nu_\mu$ and $\bar{\nu}_\mu$ have similar properties, 
$\nu_\mu$ and $\nu_\tau$ are produced by neutral currents (NC) 
in the same manner
and probably, muons are present only in the innermost core; 
thus $\nu_\mu$, $\bar{\nu}_\mu$, $\nu_\tau$ and $\bar{\nu}_\tau$
should have a very similar distribution.

Let us begin by describing the general properties of the neutrino fluxes.
As seen in Sec.\ref{ssec:nf}, in 
the delayed scenario the collapse has four main phases.
Correspondingly, we distinguish between an early neutrino emission,
during the  ``infall'' and ``flash'' phase and 
a late phase of emission (or `thermal phase'), during  the ``accretion'' and ``cooling'':
see Tab.~\ref{tab:SNcoll} for more details.
The most uncertain phase is certainly the  one of ``accretion'',
that, together with ``cooling'', 
accounts for most of the energetics.
Perhaps, one could argue that a fair estimate
of errors should be just 100 \%. 
In support of this (apparently too conservative) statement, 
we recall that we have not {\em ab initio}
calculations of these fluxes and alternative 
(even if incomplete or still speculative)
scenarios have been considered.
Furthermore, the calculations 
that tried to estimate the effect of rotation
(by Imshennik and collaborators and more 
recently by Fryer and Heger \cite{f}) found very different 
fluxes and in particular a severe suppression
of muon and tau neutrinos.\footnote{If 
three dimensional effects  have an essential role 
detectable gravitational burst can occur; this adds 
interest in carrying to fulfillment these complex simulations.}

Reference ranges on neutrino energies averaged on time (starting 
at flash time, $t_0=t_{fl}$) found comparing a number 
of numerical calculations are:
\begin{equation}
\begin{split}
\langle E_{\nu_e} \rangle=10-12 \mbox{ MeV },\\
\langle E_{\bar{\nu}_e} \rangle=11-17 \mbox{ MeV },\\
\langle E_{\nu_x} \rangle=15-25 \mbox{ MeV },
\end{split}
\label{eq:aven}
\end{equation}
The reason of this hierarchy is that neutrinos that interact
more  -- $\nu_e$ and $\bar \nu_e$ undergo CC reactions, beside NC -- decouple in more external regions 
of the star at lower temperature. In other words, each neutrino type has its own  `neutrino-sphere' -- 
$\nu_e$'s one being the outermost. \\
The approximate amount (average values over flash and accretion and cooling duration)
of the total energy ${\cal E}_B$ carried away 
by the specific flavor is 
\begin{equation}
\begin{split}
{\cal E}_{\nu_e} = f_{\nu_e}  \ {\cal E}_B ~~~\mbox{with:}  ~~f_{\nu_e} = 17-22\ \%\\ 
{\cal E}_{\bar{\nu}_e}  = f_{\bar\nu_e}  \ {\cal E}_B ~~~\mbox{with:}  ~~f_{\bar\nu_e} = 17-28\ \%\\ 
{\cal E}_{\nu_x}  = f_{\nu_x}  \ {\cal E}_B  ~~~\mbox{with:} ~~ f_{\nu_x} = 16-12\ \%
\end{split}
\label{eq:fi}
\end{equation}
\noindent The approximate equality 
found in numerical calculations
has been  called `equipartition', but in
our understanding, there is no profound 
reason behind this result.
We recall again that these numbers should be regarded with caution.\\

Next, we would like to introduce 
a general formalism to describe {\em parameterized 
neutrino fluxes}.
Such a description
requires {\em (a)}~to know the distance of production $D$,
{\em (b)}~ to assume a distribution over the solid angle 
(usually this is isotropic, up to corrections of 
the order of $\sim 1-10$ \% at most)
{\em (c)}~to assign a `luminosity' function 
$ d {\cal E}_i/dt$ {\it (energy carried by neutrinos per unit time)} 
for each neutrino species,
{\em (d)}~to describe the neutrino energy spectrum, 
presumably black-body of Fermi-Dirac type:
\begin{equation}
n(E; \eta_i, T_i)=\frac{1}{N} \frac{E^2}{[1+\exp(E/T_i - \eta_i)]}~~~ 
\mbox{ with the normalization factor} ~~N = T^3_i F_2(\eta_i)
\label{eq:FD}
\end{equation}
where 
$T_i$ is the temperature expressed in energy units
(the normalization factor $N$ and the meaning of $F_2$ 
are explained in App.~\ref{fermi}).
Possible non-thermal effects are often 
described by introducing a parameter $\eta_i$ 
that modifies the shape of the distribution.
This parameter is not a chemical potential 
(it is not subject to the condition $\eta_\nu=-\eta_{\bar{\nu}}$) 
and it is often called the pinching factor.\footnote{This name 
arises since, for fixed average
energy $\langle E\rangle$, a value
$\eta>0$ leads to a distribution suppressed
at low and high energies. The reason \cite{janka}
why this happens at high energy is simply that hotter
neutrinos are in contact with cooler 
regions than average neutrinos.
A typical cross-section that increases
fast with energy and has a large threshold is $\nu_e {}^{12}\! C\to
 {}^{12}\! N e^-$: changing $\eta$ from zero to 2 
decreases by 20 \% the event number, 
if $\langle E\rangle=23$ MeV.}  (However, it is not 
excluded that the true non-thermal effects are 
even more dramatic and the high energy tail 
of the spectrum is cutoff as $e^{-(E/E_0)^2}$, where $E_0$ 
is another new parameter).\\
At a distance $D$ from the source,  
the flux differential in energy and time ($t\ge t_{fl}$)~is:
\begin{equation}
\frac{d^2\Phi^0_i}{dE dt} = \frac{1}{N^\prime} \frac{1}{4\pi D^2} \frac{d{\cal E}_i}{dt}~ n(E; \eta_i, T_i) ~~~ 
\mbox{with the normalization factor} ~~ N^\prime = T_i \frac{F_3(\eta_i)}{F_2(\eta_i)}
 \label{eq:flux}
\end{equation}
(the index $0$ recalls us that we do not take into account
oscillations during propagation
and the  normalization factor corresponds to the average neutrino energy $N^\prime = \langle E_i\rangle$, 
as recalled in App.~\ref{fermi}). 
Therefore, one has to calculate or to reconstruct experimentally
three functions of the time for each type of neutrino, 
${\cal E}_i$, $T_i$ and $\eta_i$ which might be a difficult task.
\begin{figure}[t]
\begin{center}
\includegraphics[width=12.8cm]{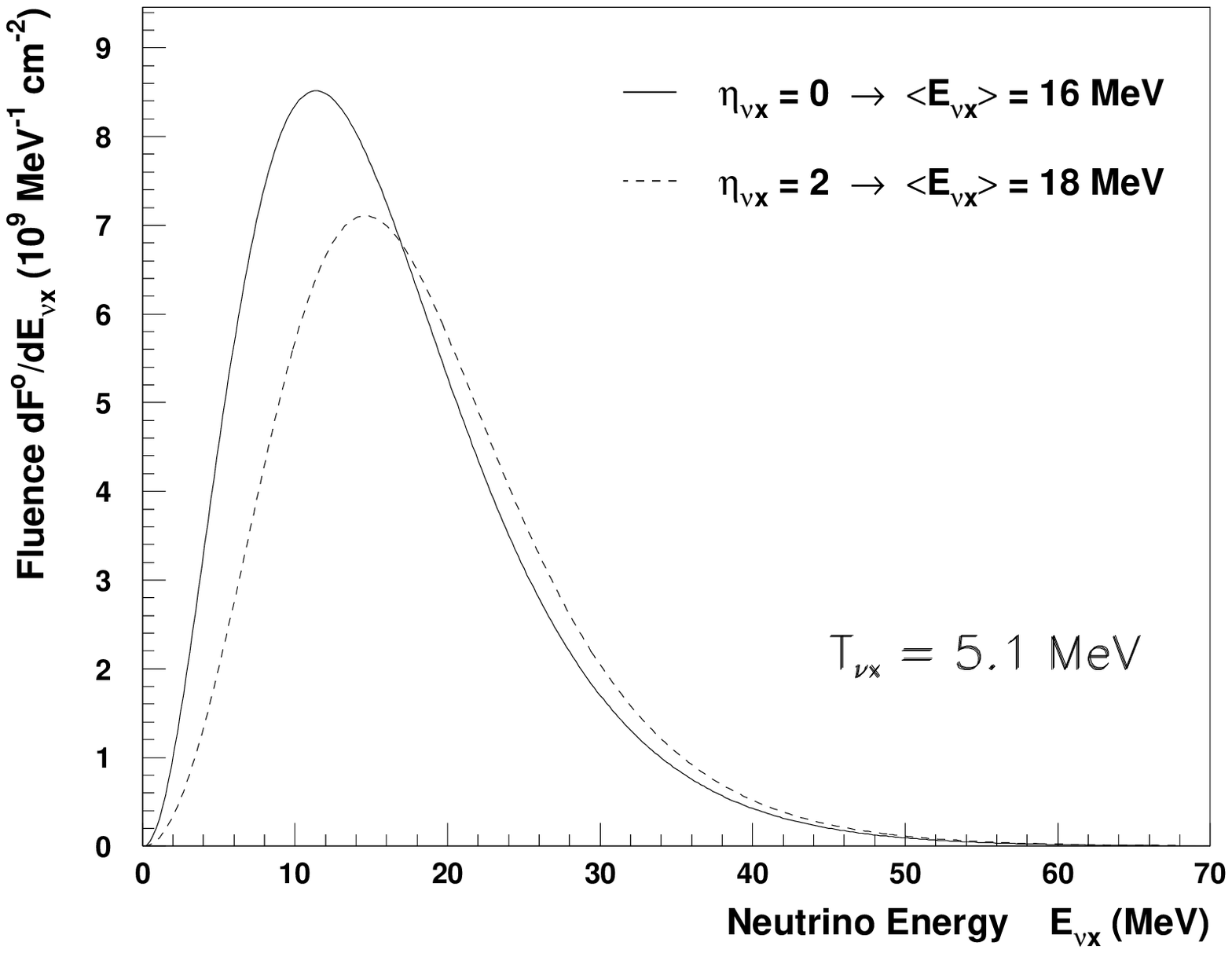}
\caption{\footnotesize\textsf{\textit{Fluence spectra 
for $\nu_x$-type neutrinos. No oscillation effect is accounted here. 
The effective temperature parameter 
is set at a reference value $T_{\nu_x}= 5.1$~{\rm MeV}. Curves refer to 
two different values of the effective 
pinching parameter $\eta = 0, 2$. The other SN parameters are 
$D=10$ kpc, $f=1/6$, ${\cal E}_B=3\cdot 10^{53}$ erg.}}}
\label{fig:fluence}
\end{center}
\end{figure}

For this reason, or just to get a `synthetic' description, 
it is common use to introduce the {\em time integrated fluxes}, i.e.~the neutrino `fluences' $F^0_i$ 
from the thermal phase, that are parameterized in a very similar manner, namely by
{\em (1)}~an energy fraction parameter $f_i$
$$
\int\limits_{t_{accr}} dt \frac{d{\cal E}_i}{dt}=(1-\delta_{fl}) f_i {\cal E}_B
$$
(here, we singled out the energy fraction $\delta_{fl} {\cal E}_B$ 
that goes in the $\nu_e$ `flash' and fractioned the rest by $f_i$), by 
{\em (2)}~an effective temperature $T_i$ (time averaged value from $t_{accr}$) that characterize the spectrum, 
and finally by 
{\em (3)}~an effective $\eta_i$ parameter  for non-thermal effects.\\
In summary, the energy differential fluence, for each neutrino species and for a distance $D$ from the source, is given by:
\begin{equation}
\label{eq:flnor}
\frac{dF^0_i}{dE} = \frac{1}{4\pi D^2} ~(1-\delta_{fl}) f_i {\cal E}_B ~\frac{1}{T^4_i F_3(\eta_i)}~
 \frac{E^2}{[1+\exp(E/T_i - \eta_i)]}
\end{equation}
Integrating the fluence over the whole surface of emission 
and over all neutrino energies
$$
{4\pi D^2} \cdot \int dE \frac{dF^0_i}{dE} = \frac{(1-\delta_{fl}) f_i {\cal E}_B}{\langle E_i \rangle}
$$
we find the number of $i$-type neutrinos ($N^0_i$) 
emitted during thermal phase (accretion and cooling) --  
oscillations not yet accounted.
In Fig.~\ref{fig:fluence} typical $\nu_x$-type 
fluence spectra described by Eq.~(\ref{eq:flnor}) are shown.

We would like to argue that a minimal set of parameters, 
beside to ${\cal E}_B$ from Eq.~(\ref{eq:eb}) and $\delta_{fl}$ 
from Tab.~\ref{tab:SNcoll}, should include
the following ones:
\begin{equation}
\mbox{$T_{\bar \nu_e}$, $\kappa=T_{\nu_x}/T_{\bar\nu_e}$, $f$ and $\eta$}
\label{eq:param}
\end{equation}
These parameters have the following meaning:
\begin{itemize}
\item $T_{\bar\nu_e}=$ the effective temperature of electron antineutrinos
(presumably, easier to observe);
\item $\kappa=$ increase in temperature of 
$\mu$/$\tau$ (anti)$\nu$
(oscillations and NC reactions imply this parameter);
\item $f=f_{\nu_e}=f_{\bar\nu_e}=$ the fraction of 
electron (anti)neutrinos, presumably $f=1/4-1/6$ (see Eq.~\ref{eq:fi}), 
which constrains $f_{\nu_x}=(1-2 f)/4$ (the case $f=1/6$
represents exact `equipartition');
\item an effective
pinching parameter $\eta\ge 0$,
equal for all types of neutrinos
(that is not expected to be accurate, but could be 
adequate in practice). 
\end{itemize}
Usually, $T_{\nu_e}$ is not a very important 
parameter to describe the neutrino signal,  
simply because this is the lowest temperature, 
however this can be estimated 
by a `reasonable' condition on the emitted lepton number
$N_{\nu_e}^0-N_{\bar\nu_e}^0$ and the parameters of 
Eq.~(\ref{eq:param}):\\[1ex]
$$ T_{\nu_e}=T_{\bar\nu_e}/
[1+  (N_{\nu_e}^0-N_{\bar\nu_e}^0) 
(T_{\bar\nu_e} F_3(\eta)/F_2(\eta))\, /\, (f {\cal E}_B)]$$

At a further level of refinement, we may
introduce  time dependent 
features and distinguish between `cooling' 
and `accretion' neutrinos.
E.g., we have a cooling component 
whose luminosity $d{\cal E}_i/dt$ scales as $T^4_i$, 
and whose temperature obeys a time law as:
$$T_i(t)=T_i(0)/(1+t/\tau)$$
(the constant $\tau\sim 10-100$ sec 
has to be extracted from the data or computed).
On top of that, we add for $t<\Delta t_{accr}$ another rather
luminous phase, presumably with 
a marked non-thermal behavior ($\eta\neq 0$)
and with its own effective temperature.
Since the efficiency of energy transfer 
to matter is not large, (anti)$\nu_e$ should carry 
a sizable  fraction of energy
($\nu_x$ are of little use to revive the shock,
but perhaps, only few of them are produced in this phase).

Sometimes, simplified models of the emission are introduced
(see e.g.~\cite{ll}). 
Most commonly, one describes the cooling phase as 
a black-body emission  from effective 
``neutrino radiation'' spheres.\footnote{Even if, one 
could believe that expected deviation from spherical symmetry are 
large, especially for early phases of neutrino emission 
and for deep layers of the 
collapsing star.}  
Similarly, one can model the accretion
phase by suggesting that the non-thermal 
neutrino production is from $e^\pm$ interactions 
with the accreting matter.
This suggests that the fluxes are proportional 
to the cross-sections: thus, their scaling should 
be more similar to $E_\nu^4$ than to $E_\nu^2$. 
It is rather interesting that there
is some hint of such a luminous phase already from
SN1987A neutrino signal, see again Ref.~\cite{ll}.\footnote{We 
would like to comment on the {\em numerical} 
estimate of \cite{ll}, that in SN87A about 20~\% of ${\cal E}_B$ was 
emitted during accretion. This is not far 
from the `standard' estimate of 10 \%
reported in Tab.~\ref{tab:SNcoll}, however 
the agreement improves further if we assume that $\nu_x$ are not emitted
during accretion (rather than equipartition). In fact, 
$N_{\bar{\nu}_e}\propto 20\ \%\ {\cal E}_B/6
\approx 0.7 \cdot 10\ \%\ {\cal E}_B/2$ (the factor 0.7 accounts
for oscillations, neglected in Ref.~\cite{ll}).}
In our view, this indication is encouraging for theory and 
for future observations, even though this 
is not supposed to convince skeptics.

\subsection{Effects of neutrino oscillations\label{ssec:no}}
The basics of three neutrino oscillations and matter enhanced conversion 
mechanism are briefly set out in App.~\ref{o}.
Here we apply them for the supernova, an environment 
characterized by very high electron and 
baryon densities and of course by 
very intense neutrino fluxes.

Let us then start by describing the effect of neutrino oscillations
in the stellar medium.
Oscillations do not affect neutral current events
if we postulate to have only 3 types of neutrinos.
In fact, the fluence $F_{e}^0+F_\mu^0+F_\tau^0$ 
is not changed by reshuffling the 
fluxes (`NC are flavor blind').
Oscillations modify only charge current  events (CC).
To describe this phenomenon,
we need just two functions $P_{ee}$ and $P_{\bar{e}\bar{e}}$, 
the electron neutrino/antineutrino survival 
probabilities, since the $\mu$ and $\tau$ flux are 
supposed to be identical.\footnote{Indeed,
we see a $\nu_e$ if it 
stays the same or if $\nu_\mu$ or $\nu_\tau$ oscillate into $\nu_e$:
$F_{\nu_e}=P_{ee} F_{\nu_e}^0+P_{\mu e} F_{\nu_\mu}^0+P_{\tau e} F_{\nu_\tau}^0$. Rewriting
$F_{\nu_e}=P_{ee} F_{\nu_e}^0+(P_{\mu e} +P_{\tau e}) F_{\nu_x}^0$ and 
recalling that $1=P_{ee}+P_{\mu e} +P_{\tau e}$, we conclude the proof.}
In order to calculate $P_{ee}$, one has to solve the
evolution equation described by the effective  hamiltonian
(see again App.~\ref{o})
\begin{equation}
H_{eff}=2.533\cdot U \frac{\mathrm{diag}(m_i^2)}{E_\nu} U^\dagger ~+~
3.868\cdot 10^{-7}\ \rho Y_e\cdot \mathrm{diag}(1,0,0)
\label{heff}
\end{equation}
where $H_{eff}$ is in m$^{-1}$,
$\nu$ masses $m_i$ are in eV and the energy $E_\nu$ is in  MeV
(similarly for $\bar{\nu}_e$, with $U\to U^*$, 
and the second term with opposite sign).
The first and second term in the r.h.s.~of Eq.~(\ref{heff}) corresponds to 
the `vacuum' term and to the `matter' term, respectively.\footnote{There 
is an additional `matter term' due to neutrino forward scattering on 
background neutrinos. Its effect on neutrino oscillations with masses 
as in App.~\ref{o} is small, but to see why one needs to 
go into subtleties. In fact, 
during the most luminous phase (the ``flash'')
the density of background neutrinos 
$N_\nu\approx d{\cal E}/dt/(\pi r^2 c \cdot  \langle E\rangle)$
is $10-100\times N_e$
around the point $r_*\sim 5\cdot 10^4$~km 
where $\theta_{13}$ gives MSW conversion.
However, the additional matter term is strongly suppressed 
in comparison with usual one, since the relevant current 
is not $N_e (1,\vec{0})$ (background electrons at rest) but rather 
$N_\nu p/E$ (relativistic background neutrinos).
When this current is contracted with the 
current of propagating neutrinos $\bar{u}(p') \gamma_a (1-\gamma_5) u(p')$, 
it gives zero up to the square of the deviation 
from collinerity between propagating 
and background neutrinos, that is $\sim (d/r_*)^2\lesssim 10^{-5}$, 
where $d$ is the dimension of the source. The new term is 
below 1 \% at $r=r_*$ and scales as $r^{-4}$; thus its effect is small.
A strictly related and more detailed discussion
is in Y.Z.~Qian and G.~Fuller, Phys.Rev.D {\bf 51} (1995) 1479.}
In the latter one the supernova density $\rho$ (in g/cm$^3$) and the 
fraction of electrons $Y_e=N_p/(N_p+N_n)$ must be taken
from some pre-supernova model. 
For orientation, a pre-supernova mantle  density 
$\rho =  100-200\ (r_0/r)^3$ g/cm$^3$ 
with $r_0=10^5$ km and $Y_e\sim 1/2$ can be used.\footnote{Note 
that we are assuming that the pre-supernova dynamics {\em does not} 
modify in an essential way the structure of the mantle of the star.
The modifications due to the shock wave, usually, do not
lead to large effects. However it is possible at least in principle 
that a massive occurrence of stellar winds, explosive nuclear reactions, 
and/or instabilities modifies the mantle before the occurrence 
of the core collapse. These possibilities will be 
better tested by astronomical observations 
of the pre-supernova, by the study of the supernova spectra and/or 
possibly by theoretical modelling of the star mantle; e.g., 
using the delay between the neutrino burst and the light.
(We thank Marco Selvi for this important remark.)}\\
Inside the core of the star, the `matter term'  dominates and the produced 
$\nu_e$ coincides with the heaviest state 
of the effective hamiltonian.
It can  happen that $\nu_e$ always coincides with the local
mass eigenstate during propagation e.g., $\nu_e\equiv \nu_3^m(t)$ -- 
this is usually called `adiabatic' conversion. 
\begin{figure}[t]
\begin{center}
\includegraphics[width=12.9cm]{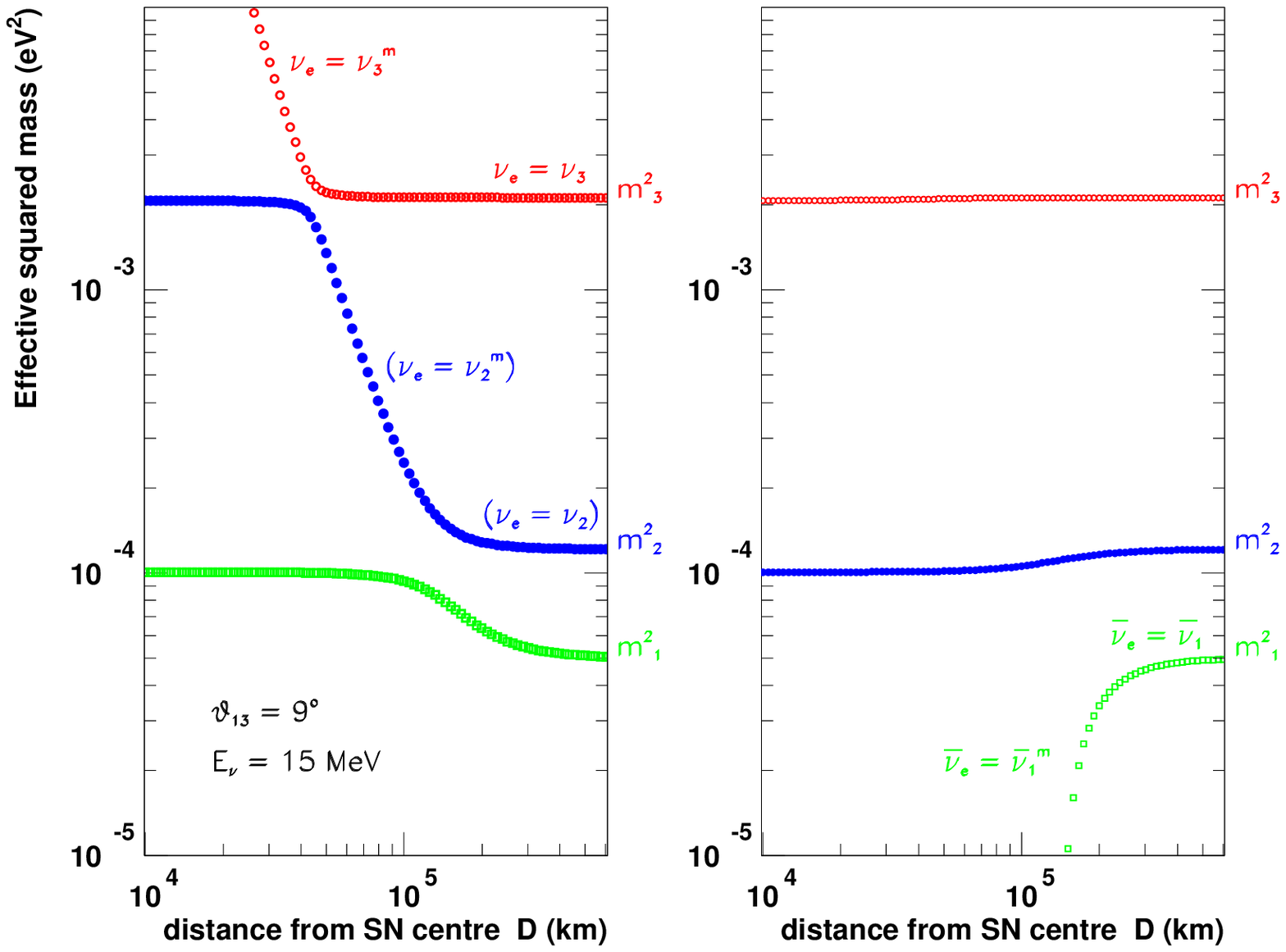}
\caption{\footnotesize\textsf{\textit{These plots show the 
effective neutrino masses 
of neutrinos [Left panel] and antineutrinos [Right panel] inside the star.
As visible from $r\to\infty$ regions, we assume 
that the neutrino spectrum 
obeys a `normal' mass hierarchy. 
(We do not emphasize another possibility compatible with the
data, `inverse' mass hierarchy. The common mass scale
is immaterial for oscillation however.)\label{fig1}}}}
\end{center}
\end{figure}
At the exit of the star, neutrinos propagate freely as 
mass eigenstate in vacuum e.g., $\nu_e\equiv \nu_3$, 
see Fig.~\ref{fig1}~[Left]. 
This depends on the unknown size of the 
vacuum mixing angle $\theta_{13}$ ($U_{e3}=\sin\theta_{13}$)
and on the electron density distribution in the 
star $N_e\propto \rho Y_e$.
The approximate values of $\theta_{13}$ when adiabatic 
conversion should occur
are shown in the following equation:
\begin{equation}
P_{ee}=|\langle \nu_e|\nu_e(t)\rangle |^2=\left\{
\begin{array}{lcl}
U_{e3}^2\sim 0 & \mbox{ if }\theta_{13}>1^\circ & \mbox{adiabatic conversion of }
\nu_e\to\nu_3 \\
U_{e2}^2\sim 0.3 & ~~\mbox{ if }\theta_{13}<0.1^\circ & \mbox{adiabatic conversion of }
\nu_e\to\nu_2 
\end{array}
\right.
\end{equation}
(at present, we cannot exclude 
that $\theta_{13}$ falls in an intermediate case).\footnote{This qualitative discussion of 
neutrino oscillations in matter as illustrated  
in Fig.~\ref{fig1} corresponds to the 
approximated analytical expression for the probability of survival
$P_{ee}=U_{e2}^2 P_H+U_{e3}^2 (1-P_H)$, where the 
`flip probability' associated with $\theta_{13}$ is 
$P_H=\exp[ - U_{e3}^2/(4\cdot 10^{-5})\cdot  
(20\mbox{ MeV}/E_\nu)^{2/3}]$. 
(The analytical expression of the exponent is 
$2\pi r_0 (\sqrt{2} G_F N_{e0})^{1/3} (\Delta m^2/2 E)^{2/3}$
and assumes that $N_e(r)=N_{e0}\; (r_0/r)^3$, see last reference
in \cite{msw}).
The corresponding flip probability associated 
with solar mixing is $P_L=0$ with 
good approximation.}
Solar neutrino mixing makes almost certainly adiabatic 
the second conversion, if the first should fail.\\ 
Thus, accounting fo oscillations, the fluence of $\nu_e$
becomes:
\begin{equation}
F_{\nu_{e}}=
\left\{
\begin{array}{lc}
F_{\nu_x}^0 & \mbox{ if }\theta_{13}>1^\circ \\
0.3 F_{\nu_e}^0+0.7 F_{\nu_x}^0 &
~~\mbox{ if }\theta_{13}<0.1^\circ 
\end{array}
\right.
\label{eq:nuconv}
\end{equation}
Similarly, for antineutrinos 
$\bar{\nu}_e=\bar{\nu}_1^{m}(t)$, see Fig.~\ref{fig1} [Right], that implies
$P_{\bar{e}\bar{e}}=|\langle \bar{\nu}_e|\bar{\nu}_e(t)\rangle |^2=
U_{e1}^2\sim 0.7$. Thus, the formula for the flux becomes 
 $F_{\bar\nu_{e}}=0.7 F_{\bar\nu_{e}}^0+0.3 F_{\nu_x}^0$.
Now we can make the argument for oscillations: Since we expect that 
$F_{\nu_e}^0\neq F_{\nu_x}^0$ and  $F_{\bar\nu_{e}}^0\neq F_{\nu_x}^0$, 
oscillations should modify the expected supernova neutrinos fluxes.
These modifications are large 
(e.g., the flash yields little in CC:
NC events range from 70 to 100 \%)
and can be observable, but the 
message that we want to stress here is simply 
that {\em these effects should be taken into account}
in order to interpret the SN neutrino signal correctly.\\

Finally, we consider the `earth matter effect', 
possible operative if SN neutrinos cross the earth 
before hitting the detector.
We will show that, with the current oscillation parameters, 
it is not very large. As we saw, in a possible scenario
(=normal mass hierarchy, very small $\theta_{13}$) 
neutrinos exit from the star 
as $| \nu_e \rangle \to | \nu_2 \rangle $ and 
$| \bar{\nu}_e \rangle \to | \bar{\nu}_1 \rangle $ due to the 
MSW effect \cite{msw}, or in other words,
$$
P_{ee}=\sin^2\theta_{12}\sim 0.3 
~~~\mbox{ and }~~~P_{\bar{e}\bar{e}}=\cos^2\theta_{12}\sim 0.7
$$
If (anti)neutrinos cross the earth in the last stage of their path,
new oscillations will occur
(since vacuum eigenstates are 
not eigenstates in the earth matter)
and previous expressions will be 
modified. For constant density (say, earth mantle - $\rho_{\oplus}\approx 4$ 
g/cm$^3$) the 
solution of a two-flavor version of Eq.~(\ref{heff}) gives:
\begin{equation}
P_{ee}=\sin^2\theta_{12}
\left[
1+\frac{4\varepsilon \cos^2\theta_{12}}{
(1+\varepsilon)^2-4 \varepsilon\cos^2\theta_{12}
}\cdot
\sin^2\left( \frac{\Delta m^2_{12} L}{4 E} 
\sqrt{(1+\varepsilon)^2-4 \varepsilon\cos^2\theta_{12}}
\right)
\right]
\label{eme}
\end{equation}
with $\theta_{12}\approx 33^\circ$ and $\Delta m^2_{12}\sim 7\cdot 10^{-5}$ eV$^2$, 
where 
$$
\varepsilon=
\frac{\sqrt{2} G_F N_{e\oplus}}{\Delta m^2_{12}/2 E}
\simeq 9\ \% \ 
\frac{\rho_{\oplus}/(\mbox{4\ g/cm}^3)\cdot Y_{e\oplus}/(\mbox{0.5})
\cdot E/(20\mbox{ MeV})}{\Delta m^2_{12}/
(7\cdot 10^{-5}\mbox{ eV}^2)}
$$
For $\bar{\nu}_e$, just replace $\theta_{12}\to 90^\circ-\theta_{12}$.
Earth matter effect is larger than for solar neutrinos,
simply because supernova energies are larger, see Eq.~(\ref{heff}).
This can give rise to spectacular wiggles, especially if 
large energies events are seen.
Numerical considerations 
based on previous formulae suggest that this investigation will 
be demanding. If (or when) the position of the supernova 
will be known, it will be possible to include such an effect, 
reducing ambiguities in the interpretation of the signal.

\subsection{Importance of electron neutrino signal 
($\nu_e$ absorption on Argon)\label{ssec:eos}}
In order to complete the discussion and 
to show an application of the formalism, we 
will consider in detail the specific supernova neutrino 
signal provided by the reaction of absorption 
\begin{equation}
\nu_e + \mbox{Ar}\to \mbox{K$^{*}$} + e^-
\label{eq:absreac}
\end{equation}
that has a large cross-section. The 
signature for reaction (\ref{eq:absreac})
is given by a leading electron accompanied by soft electrons from 
conversion of K$^*$ 
de-excitation $\gamma$'s in the Argon volume surrounding 
the interaction vertex.
This signal could be seen by the forthcoming 
detector ICARUS \cite{icarus} based on
the liquid Argon technology.\footnote{In this discussion, 
we want to emphasize the potential of an ideal $\nu_e$ detector,
putting aside technical limitations like need of a 
long term stability of operation, 
detection threshold, efficiency and finite resolution.
Let us recall that also other detectors 
can see the $\nu_e$ signal with other reactions, even 
if usually this is not the main signal.
For instance: 
{\em (1)}~$\nu_e + {}^{16}{\rm O} \to e + {}^{16}{\rm F}$ (with $Q=15.4$~MeV)
can be exploited at water \v{C}erenkov 
detectors as Super-Kamiokande or SNO due to the angular distribution
(${}^{16}{\rm F}$ rapidly decays by proton emission); 
{\em (2)}~$\nu_e + {}^{12}{\rm C} \to e + {}^{12}{\rm N}$ (with $Q=17.4$~MeV)
can be seen in scintillators detectors (LVD, Borexino, KamLAND, BAKSAN),
with the great advantage of offering a 
double tag, due to the $\beta^+$ decay  of Nitrogen; 
{\em (3)}~$\nu_e + {\rm D} \to e + p +p$ (with $Q=1.4$ MeV and a large 
cross-section) can be used at the inner part of  
SNO (the signal is given by a lone electron, in contrast with 
neutral current, or electron antineutrino reactions on deuterium
that are tagged by additional neutrons).
$\nu_e$ detection profits of staying closer 
to the philosophy of solar neutrinos and 
of employing literally solar neutrino detectors. 
Note that a big $Q$ value or a rapid rise of the cross section 
amplifies the difference between the case with and without 
oscillations.}
(We are not going to discuss
the more difficult and important question of `what we can 
learn from supernova neutrinos', whose answer will of course depend on which 
neutrino detectors will be working when  next galactic supernova 
will explode and what will be the distance of this supernova;
but it is almost from granted that we will learn a lot from the
$\overline{\nu}_e p\to n e^+$ reaction for the reasons recalled 
in App.~\ref{xsec}). 

In a 3~kton liquid Argon detector {\bf the number of $\nu_e$-absorption events
is about 400}, for a supernova exploding at $D=10$ kpc. 
To calculate this number, one simply multiplies 
the fluence (including oscillations) 
by the number of target nuclei and by 
the cross-section of the reaction, 
and than integrates over the  possible 
neutrino energies. 
In the present calculation we employed a `hybrid model' 
for the cross-section of the reaction in Eq.~(\ref{eq:absreac}):
shell model for allowed transitions, 
and random phase approximation 
(RPA) for forbidden ones (see below). The other inputs were:
{\em (a)}~a normal hierarchy of neutrino masses; 
{\em (b)}~$\theta_{13}$ large enough to produce $P_{ee} \sim 0$ -- that 
    is, $\nu_e\to \nu_3$  producing $F_{\nu_e}\equiv F^0_{\nu_x}$;
{\em (c)}~an exact equipartition of the fluxes ($f={1}/{6}$) 
    and ${\cal E}_B = 3\times 10^{53} \mbox{ erg}$;
{\em (d)}~a spectrum without pinching ($\eta = 0$);
{\em (e)}~a neutrino temperature of  $T_{\nu_x}= 5.1$~MeV, 
   corresponding to an average energy $\langle E_{\nu_x}\rangle = 16$ MeV
(following the indications of the most 
recent calculations \cite{delayed}, we 
assume that in absence of oscillations the 
temperature of electron neutrinos is $T_{\nu_e}=3.5$ MeV, 
that is closer to $T_{\nu_x}$ than thought in the past).

The use of adequate cross-section for neutrino absorption 
reaction on Argon is important.
 Allowed transitions to low lying Potassium (K) excited 
 levels\footnote{The allowed transitions in  
Eq.~(\ref{eq:absreac}) include 
two contributions: {\em (1)} Fermi
 transitions from $^{40}$Ar ($J^\pi=0^+, T=2$) to the isobaric 
 analog state of $^{40}$K ($J^\pi=0^+, T=2$) 
 at an excitation energy of 4.38 MeV and
 {\em (2)}~Gamow-Teller transitions to several low lying $J^\pi=1^+, 
 T=1$ states of $^{40}$K with excitation energies
 between 2.29 to 4.79 MeV.}  dominate for neutrino energies less than 
 $\sim 15$~MeV (i.e.\ in the energy range of interest for solar 
 neutrino experiment). 
 Shell model computation  \cite{ormand} allows to reliably describe 
 the allowed cross-section. 
 At higher energies, as for the SN case here considered, 
 forbidden transitions become relevant as well. These are dominated 
 by the collective response to giant resonance,
 so that the RPA model \cite{langanke}   is usually considered sufficient to 
 describe the non-allowed contributions
 to the ($\nu_e$, Ar) cross-section.\footnote{In 
the RPA calculation of Ref.~\cite{langanke},
all forbidden transitions 
to $^{40}$K levels with $J\le 6$ and both parities have been
included.} 
\begin{figure}[t]
\begin{center}
\includegraphics[width=12.9cm]{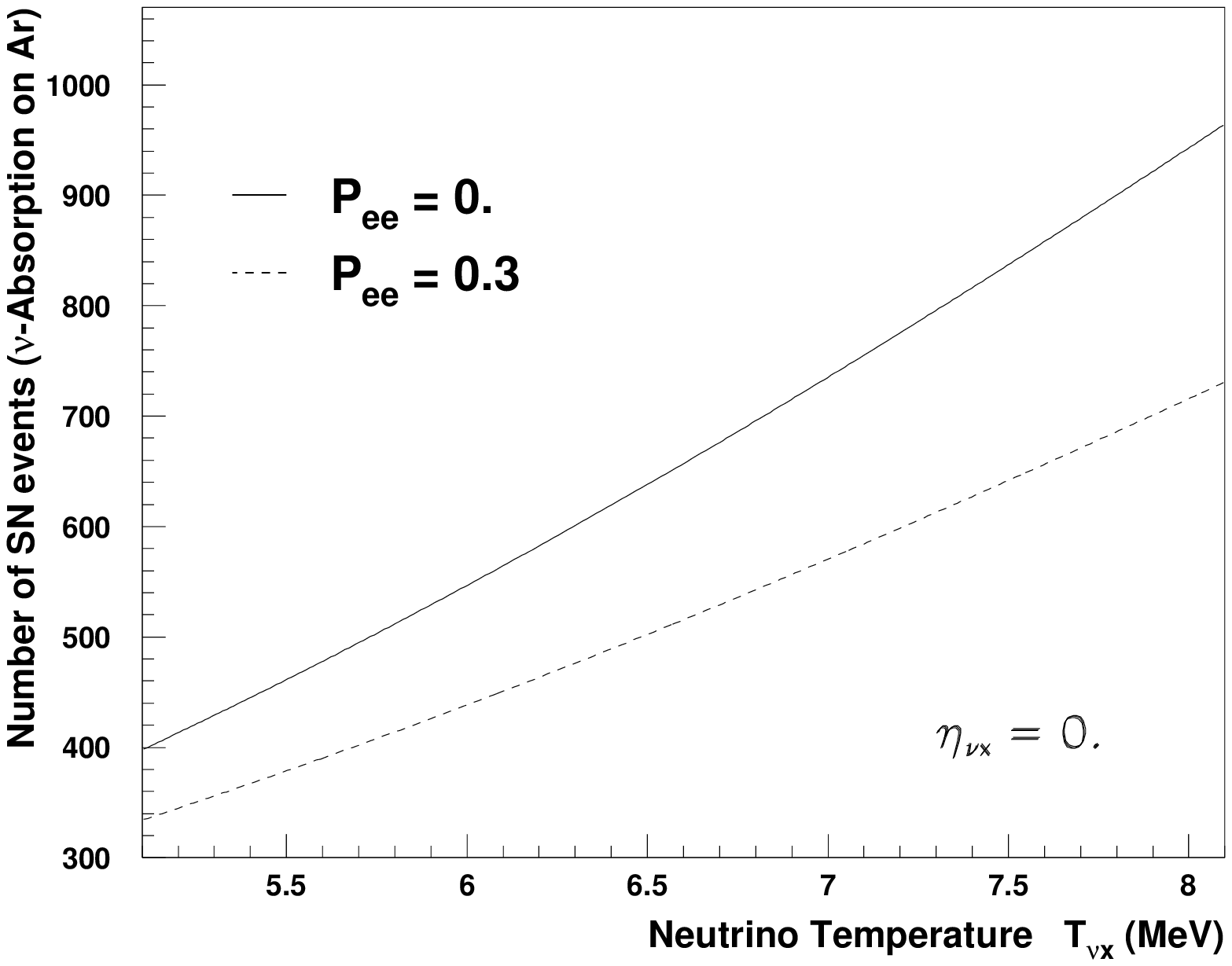}
\caption{\footnotesize\textsf{\textit{Number of expected $\nu_e$ 
events as a function of the effective temperature.
Oscillation are separately accounted for - full line and dotted line - 
according to the two reference cases of Eq.~(\ref{eq:nuconv}).
Neutrino fluence is taken from Eq.~(\ref{eq:flnor}), with $T$ free to vary 
and pinching parameter set at $\eta= 0$.
The other SN parameters are: $D=50$~kpc, $f=1/6$ (strict equipartition), 
and ${\cal E}_B = 3\times 10^{53}$~erg.
The cross-section `hybrid model'  for $\nu_e$-absorption on Argon  
is used. A reference 3~kton detector mass (corresponding to
$3\times 1.5\cdot 10^{31}$ Ar targets) is considered. 
For simplicity, we assume an ideal detector, 
without threshold on the final state electron energy
and full detection efficiency.}}}
\label{fig:nevtemp}
\end{center}
\end{figure}

How the number of events changes, with reasonable changes 
of the input parameters? To answer this question, we 
can calculate 
the percentage variation $100\times \delta N/N$
of the number of absorbed $\nu_e$  
under a number of alternative hypotheses:
\begin{center}
\begin{tabular}{ c|c|c|c|c}
${T+\Delta T}$     &  ${T-\Delta T}$ &   ${f\to {1}/{8} }$  &   ${\eta\to 2}$ &  ${P_{ee} \to 0.3}$   \\
\hline
$+51$ \%     & ${- 45}$ \%   &   ${-25}$ \% & ${+15}$ \%  & ${- 16}$ \%  \\
\end{tabular} 
\end{center}
The first two columns show the effect of 
changing the temperature by $\Delta T=\pm 1.3$ MeV;
the third column, describes the effect of non-equipartitioned fluxes;
the fourth one, the effect of having a pinched (`non-thermal') spectrum;
the last column, assumes that $\nu_e\to\nu_2$ due to very small 
$\theta_{13}$.
This shows that the present uncertainty in the 
temperature has a big impact on the expected signal, about $50$~\%.
It shows also that a {\em mixture} of 
various phenomena can affect the flux at the $\sim 20$~\% level. 
To separate these effects clearly, it will be important to study several 
properties of the neutrino signal, like distributions 
in time and energy and use  several reactions.
In Fig.~\ref{fig:nevtemp} we show the calculated 
number of expected events for a wide range of values  of 
the effective temperature.

In some situations, the electron neutrino signal
can lead to `model independent' inferences. 
For instance, if it were possible to demonstrate that the earth 
matter effect (associated with solar 
$\Delta m^2\approx 7.1\cdot 10^{-5}$ eV$^2$)
occurs in $\nu_e$ events {\em and} in $\bar{\nu}_e$ events, 
we would have a proof that $\theta_{13}$ is small.
If instead it occurs {\em only} for $\bar{\nu}_e$, the converse is true
and furthermore, the hierarchy must be normal
(there is an adiabatic conversion associated with 
the heaviest neutrino). It should be remarked 
however that a `golden' observation (that is seeing 
one or more wiggles) requires a great precision in energy measurement 
or a lucky configuration, 
namely, a supernova exploding just below the horizon.
In fact, the phase of oscillation with solar 
$\Delta m^2$ is close to $\pi/2$ for lengths of propagation 
through the earth 
of the order of $350 \mbox{ km}\times E_\nu/(20\mbox{MeV})$,
see Eq.~(\ref{eme}).

But note that even the absence of a $\nu_e$ signal would be a precious 
information. Indeed, to help the `delayed explosion' to take place,
it would be better to have a depletion of $\nu_x$ during accretion. In 
that case, the number of $\nu_e$ events in the first half-a-second 
should be small, due to Eq.~(\ref{eq:nuconv}), whereas they should be seen
during cooling.
Similarly, if non-standard scenarios (like 
collapse with rotation) are realized, $\nu_x$ 
can be depleted also during  
the cooling phase. In this case, $\nu_e$ events would be 
rare even during cooling.

\section{Summary and discussion\label{sec:sum}}
In this introduction to $\nu$-astronomy, 
we focused mostly on supernova neutrinos. 
We aimed at helping the orientation of a reader in this field,
so we did not attempt to give a comprehensive study (i.e.\
we did not consider all theoretical possibilities or 
scenarios, or reactions to detect neutrinos).
Rather, we offered a selection of the background information, 
provided some few formulae, reference numbers,  
and showed illustrative calculations.
Let us conclude by recalling some of the 
important points we touched:
\vskip2mm

\noindent $\star$ Neutrino astronomy is theoretically appealing 
and rich of promises. Supernova neutrinos are a very well defined 
and interesting possibility.

\noindent $\star$ 
Neutrino observations from SN1987A are not in contradiction
with the general theoretical picture. However,  
supernova explosions are still mysterious, 
and this warrants more discussion and stimulates more efforts.

\noindent $\star$  Next galactic supernova will permit us much
more precise observations and this will be certainly 
very helpful to progress. 
In particular, the response from new generation neutrino detector(s),
sensitive to various types of neutrinos and reactions 
could be of major importance (we discussed in some detail
the case of a detector like ICARUS, that combines a large mass 
with a high resolution and detection efficiency).

\noindent $\star$ The effects of oscillations are important
and have to be included.
Conversely, one could combine experiments and
use theoretical information in order to attempt 
to make inferences on oscillations, 
but astrophysical uncertainties should 
be thought as an essential systematics for this purpose.
(In other words, there are
chances to learn something on neutrinos, but, 
in our view, the primary aim of these observations is just
supernova astrophysics.)  

\noindent $\star$ All this is fine; the most 
important task left is an exercise of patience, 
$0-100$ years for next galactic supernova.

\vskip1cm

\section*{Acknowledgments}
We thank 
V.~Berezinsky, 
G.~Di Carlo, 
W.~Fulgione, 
P.~Galeotti,
P.L.~Ghia, 
D.~Grasso, 
A.~Ianni,
M.~Selvi 
and 
A.~Strumia 
for collaboration on these topics, 
E.~Kolbe,
K.~Langanke 
and 
G.~Martinez-Pinedo
for providing us with the cross-sections 
used in our numerical calculations
and 
G.~Battistoni, 
P.~Desiati, 
H.T.~Fryer,
V.S.~Imshennik,
O.G.~Ryazhskaya,
G.~Senjanovi\'c and 
M.~Turatto  
for useful discussions 
during the preparation of these notes.



\appendix
\section*{Appendices}

\section{An example of cross-section (inverse $\beta$ decay)\label{xsec}}
\def\theequation{\thesection.\arabic{equation}}
\setcounter{equation}{0}
The `inverse $\beta$-decay' reaction $\bar{\nu}_e+p\to e^+ + n$ 
is particularly important for actual neutrino detection.
Indeed, it has a large cross-section and 
water \v{C}erenkov and scintillator 
based detectors  have many free protons.
For illustration, we recall here a simple 
approximation of this reaction (from last paper of Ref.~\cite{ibd})
and refer to \cite{ibd} for a more complete discussion.

The {\bf tree level} 
cross-section in terms of Mandelstam invariants $s,t,u$ is:
\begin{equation}
\frac{d\sigma}{dt}=\frac{G_F^2 \cos^2\! \theta_C}{2\pi (s-m_p^2)}~
[ A(t)-B(t) (s-u) + C(t) (s-u)^2 ]
\label{a1}
\end{equation}
$A,B,C$ are well approximated as $c_1+t\cdot c_2$
at the energy of supernova neutrino detection:
\begin{equation}
\begin{array}{l}
A\approx M^2 (1-g^2)(t-m_e^2)-M^2\Delta^2 (1+g^2) -2 m_e^2 M \Delta g (1+\xi)\\
B\approx g (1+\xi) t \\ 
C\approx (1+g^2)/4
\end{array}
\end{equation}
where $M=(m_n+m_p)/2$, $\Delta =m_n-m_p$, $\xi=3.706$ and 
$g=-1.270\pm 0.003$.
Eq.~(\ref{a1}) is 
related by Jacobians 
to the cross-sections differential in the lepton energy 
$E_e$, or  in the angle $\theta=0-\pi$ between the incoming neutrino 
and the charged positron:
\begin{equation}
\frac{d\sigma}{d E_e}= {2 m_p} \frac{d\sigma}{dt}~~~\mbox{ and }~~~ 
\frac{d\sigma}{d \cos\theta}= \frac{\epsilon p_e}{1+
\epsilon(1 - \frac{E_e}{p_e}\cos\theta)} \frac{d\sigma}{d E_e}
\end{equation}
Of course, for the first formula
one has to express the Mandelstam variables in terms
of $E_\nu$ and $E_e$, e.g.,  $t=m_n^2-m_p^2-2 m_p (E_\nu-E_e)$.
To evaluate the second formula, 
one first defines $\epsilon=E_\nu/m_p$ and
calculates the positron energy $E_e$ (and momentum $p_e$) 
from 
$E_e=[ (E_\nu-\delta)(1+\epsilon) + \epsilon \cos\theta 
((E_\nu-\delta)^2-m_e^2 k^2)^{1/2}]/k$, where 
$\delta=(m_n^2-m_p^2-m_e^2)/(2 m_p)$ and 
$k=(1+\epsilon)^2-(\epsilon \cos\theta)^2$. Note that 
$E_\nu$ is one-to-one with $E_e$ at zeroth order in $\epsilon$. 
The threshold of the reaction is at $E_\nu>1.806$ MeV.
\vspace{0.1cm}

\section{Fermi integrals and polylog\label{fermi}}
\def\theequation{\thesection.\arabic{equation}}
\setcounter{equation}{0}
Let us consider the function $f_n(x,\eta)$ where 
$n=1,2,3...$ and $\eta$ is a real parameter:
$$
f_n(x, \eta) = \frac{x^n}{1+e^{x-\eta}}~~~\mbox{with}~~x\ge 0
$$
This function is needed to define 
the {\bf Fermi integral} of $n$-th order as follows:
\begin{equation}
F_n(\eta)\equiv \int^\infty_0 f_n(x, \eta)~ dx
=-{\rm Li}_{n+1}(-e^{\eta}) \cdot n !
\label{eqn:B2}
\end{equation}
The last expression involves  
the {\bf polylogarithm} function ${\rm Li}_n(x)$.

\noindent $\bullet$ This integral appears commonly when using the 
Fermi-Dirac distribution, that can be written 
as $T^2 f_2(x,\eta)$, with $x=E/T$.
E.g., integrating this distribution 
for all values of the energy $E$ we get 
the normalization factor of 
Eq.~(\ref{eq:FD}) of Sec.\ref{ssec:nf}:
\begin{equation}
\int^\infty_0 T^2 f_{2}(x, \eta)~dE~=~T^3~F_2(\eta)
\label{eqn:B3}
\end{equation}

\noindent $\bullet$ Eq.~(\ref{eqn:B2}) is 
also useful to express energy momenta:
\begin{equation}
\langle E\rangle =T \cdot \frac{F_3(\eta)}{F_2(\eta)},~~\ \ 
\langle E^2\rangle =T^2 \cdot \frac{F_4(\eta)}{F_2(\eta)},~~\ \ 
\langle \delta E\rangle \equiv\sqrt{\langle E^2\rangle-\langle E\rangle^2}
\end{equation}
The {\em variance} $\langle \delta E\rangle$
stays constant at better than 0.2 \% for $\eta<5$ at the value
$\langle \delta E \rangle=1.73\cdot T$. 
A useful approximate expression for the {\em average energy} 
$\langle \delta E\rangle$
can be obtained using ${\rm Li}_4(-1+y)/{\rm Li}_3(-1+y)-1
\approx (50.5 - 41.8 y - 6.4 y^2  - 2.3 y^3)/1000$,
where we have in mind the identification $y\equiv 1-\exp(-\eta)$.

\noindent $\bullet$ For $\eta=0,1,2$ we get $F_2=1.803,4.328,9.513$ and 
$F_3=5.682,14.39,34.30$.

\noindent $\bullet$ The series 
expansion ${\rm Li}_n(z)=\sum_{m=1}^\infty z^m/m^n$ 
of the polylog leads to some identities:
\begin{equation}
{\rm Li}_0(z)= z/(1-z);\ \ ~~~
{\rm Li}_1(z)= -\log(1-z);\ \ ~~~
{\rm Li}_{n+1}(z)= \int_0^z {\rm Li}_n(\tau) 
\frac{d\tau}{\tau}\ \mbox{with }z>0
\end{equation}

\noindent $\bullet$ At $z=\pm 1$, the polylogarithm  
can be expressed by the {\bf Z-function} (See Eq.~(\ref{eqn:B2})):
\begin{equation}
\begin{array}{lr}
{\rm Li}_n(1)= Z_n,& (Z_{2,3,4,5}=1.645,1.202,1.082,1.037)\\
{\rm Li}_n(-1)=-(1-2^{1-n}) Z_n &
\qquad\qquad\qquad \mbox{(connection of Bose and Fermi integrals)}
\end{array}
\end{equation}
\vspace{0.1cm}

\section{A reminder on neutrino masses and oscillations \label{o}}
\def\theequation{\thesection.\arabic{equation}}
\setcounter{equation}{0}

{\em (1)} The mixing matrix $U$ (introduced by Sakata 
and collaborators in 1962)
connects neutrino {\em fields} of given flavor and of given mass: 
\begin{equation}
\nu_\ell(x)=U_{\ell i}\ \nu_i(x), ~~\mbox{ where }\ell=e,\mu,\tau\mbox{ and }
i=1,2,3
\end{equation}
This implies a relation between 
{\em states} of ultrarelativistic neutrinos.
In fact, the decomposition in oscillators
$\nu(x)=\sum_{p\lambda}(a_{p\lambda}  u_{p\lambda}  e^{ipx} +
b_{p\lambda}^\dagger  v_{p\lambda}  e^{-ipx})$ implies
$b_{\ell}^\dagger= U_{\ell i} b_i^\dagger$ and 
$a_{\ell}^\dagger= U_{\ell i}^* a_i^\dagger$, so:
\begin{equation}
\begin{array}{l}
|\nu_\ell\rangle =U_{\ell i}^* |\nu_i\rangle \\
|\bar{\nu}_\ell\rangle =U_{\ell i} |\bar{\nu}_i\rangle 
\end{array}
\end{equation}
No change if the type of mass is Dirac instead than Majorana
(last one being theorist's favorite).

\noindent {\em (2)} From previous considerations, it follows that if we produce
a state of flavor $\ell$ at $t=0$ it will acquire overlap with other 
states at later time (``appearance'' of a new flavor) and at the 
same time it will loose overlap with itself (``disappearance''). 
This was shown by B.Pontecorvo in 1967, though,
the first idea dates back to 1957. Thus, a state with momentum $p$ becomes
\begin{equation}
|\nu_\ell(t)\rangle=U_{\ell i}^* |\nu_i (t)\rangle,
~~~\mbox{ where }|\nu_i (t)\rangle = e^{i (p x- E_i t)}\ |\nu_i (0)\rangle
\end{equation}
The energy of neutrinos with different masses 
cannot remain the same in the course of the propagation, since 
$E_i\approx p + m_i^2/(2 p)$
(ultrarelativistic approximation always applies 
to the cases of interest).
When the distance between production
and detection satisfies $L\approx t \gg E/\Delta m_{ij}^2$,
$|\nu_\ell(t)\rangle$ becomes different from $|\nu_\ell(0)\rangle$, 
if the mixings $U_{\ell i}$ are large enough.
As usual, $\Delta m^2_{ij}=m_j^2-m_i^2$.

\noindent {\em (3)} The effective hamiltonian of 
propagation in vacuum is $U{\rm diag}(m_i^2)U^\dagger/2E$   
(for antineutrinos, $U\to U^*$), but 
in matter there is an additional term 
$\pm \sqrt{2} G_F N_e{\rm diag}(1,0,0)$ 
($+$  is for $\nu_e$, $-$ for $\bar{\nu}_e$;
$N_e=\rho Y_e/m_n=e^-$ 
number density).
This term is linear in the Fermi coupling.
It describes a {\em coherent} interaction of neutrinos 
with the matter where it propagates.
This can drive neutrinos to be 
 ``local'' mass eigenstates during the 
propagation, thus exiting from a 
star as vacuum eigenstates, i.e.:  
$|\nu_e\rangle \to |\nu_2\rangle$ or $|\nu_e\rangle \to |\nu_3\rangle$. 
In the sun, this effect (named after MSW after Wolfenstein, Mikheyev
and Smirnov \cite{msw}) is partial
and it is pronounced for highest energy neutrino events, 
e.g., the CC events at SNO. In the supernova, it can be complete.

\noindent {\em (4)} Putting aside the indications 
of LSND indication (that will be tested at MiniBooNE) we 
know from a number of  experiments that the usual 3 neutrino 
flavors most likely oscillate among them
and this points to the following 
(roughly 1 sigma) ranges of the parameters \cite{datas}:
\begin{equation}
\begin{array}{lll}
\Delta m^2_{12}=7.1\pm 0.7\cdot 10^{-5}\mbox{ eV}^2;~~~
& \theta_{12}=33^\circ\pm 2^\circ & \\
\Delta m^2_{23}=2.0\pm 0.4\cdot 10^{-3}\mbox{ eV}^2; ~~~
& \theta_{23}=45^\circ\pm 7^\circ;~~ &
\theta_{13}< 9^\circ
\end{array}
\end{equation}
The 3 mixing angles given above parameterize 
the unitary mixing matrix $U_{\ell i}$:
$$
|U_{e3}|=\sin\theta_{13},\ \ 
|U_{e2}/U_{e1}|=\tan\theta_{12}\ (\mbox{solar mix.}),\ \ 
|U_{\mu 3}/U_{\tau 3}|=\tan\theta_{23}\ (\mbox{atmospheric mix.}).
$$
\noindent {\em (5)} We have some bounds on neutrino masses 
(in this sense, we know something more than the 
oscillation parameters $\Delta m^2_{ji}$)
from other sources: 
from $\beta$-decay (Mainz, Troitsk), 
$\sqrt{\sum_i |U_{ei}^2| m_i^2}\le 2.2$~eV;
from neutrinoless double beta decay 
(Heidelberg-Moscow at Gran Sasso, IGEX), 
$|\sum_i U_{ei}^2 m_i|\le 0.3-1$ eV
[a claim was made that the transition 
has been observed, but in our opinion, 
with a weak significance];
from galaxy surveys (2dF) 
$\sum_i m_i<1.8$ eV
or combined cosmological results (including WMAP)
$\sum_i m_i<0.7$ eV.  
We do not discuss them further, 
but we note that they suggest that the kinematic
search of effects of neutrino masses 
with SN neutrinos is difficult or
impossible.
\newpage
\footnotesize 
\frenchspacing
\begin{multicols}{2}

\end{multicols}

\end{document}